\documentclass[12pt]{article}
\usepackage{amsmath,amssymb,amsfonts,graphicx,geometry,bm}
\usepackage{tikz}
\usepackage{amsmath}
\usetikzlibrary{positioning,arrows.meta,decorations.pathreplacing}

\title{Quantum Circuit Reasoning Models: A Variational Framework for Differentiable Logical Inference}
\author{Andrew Kiruluta \\ UC Berkeley, School of Information}
\date{\today}

\begin{document}
\maketitle

\begin{abstract}
This report introduces a novel class of reasoning architectures, termed  Quantum Circuit Reasoning Models (QCRM), which extend the concept of Variational Quantum Circuits (VQC) from energy minimization and classification tasks to structured logical inference and reasoning. We posit that fundamental quantum mechanical operations, superposition, entanglement, interference, and measurement, naturally map to essential reasoning primitives such as hypothesis branching, constraint propagation, consistency enforcement, and decision making. The resulting framework combines quantum-inspired computation with differentiable optimization, enabling reasoning to emerge as a process of amplitude evolution and interference-driven selection of self-consistent states. We develop the mathematical foundation of QCRM, define its parameterized circuit architecture, and show how logical rules can be encoded as unitary transformations over proposition-qubit states. We further formalize a training objective grounded in classical gradient descent over circuit parameters and discuss simulation-based implementations on classical hardware. Finally, we propose the  Quantum Reasoning Layer (QRL)  as a differentiable hybrid component for composable reasoning models applicable to scientific, biomedical, and chemical inference domains.
\end{abstract}

\section{Introduction}

The concept of \textbf{Variational Quantum Circuits (VQCs)} has become central to the emerging field of quantum machine learning, serving as a bridge between the continuous evolution of quantum states and the optimization frameworks of classical computation. A VQC is essentially a quantum circuit with tunable parameters---rotation angles or coupling constants---that are adjusted using a classical optimizer to minimize a loss function. This variational hybrid paradigm was first popularized in the context of quantum chemistry through the \textit{Variational Quantum Eigensolver (VQE)} proposed by Peruzzo et al.~\cite{peruzzo2014variational}, where quantum hardware is used to prepare parameterized quantum states and classical algorithms iteratively update parameters to minimize the expected value of the system’s Hamiltonian. Subsequent developments such as the \textit{Quantum Approximate Optimization Algorithm (QAOA)} by Farhi et al.~\cite{farhi2014quantum} extended this approach to combinatorial optimization, while Schuld and Killoran~\cite{schuld2019quantum} and Havlíček et al.~\cite{havlivcek2019supervised} demonstrated that VQCs can serve as powerful quantum feature maps and classifiers, drawing direct analogies to kernel methods in machine learning.

The historical motivation for these developments lies in a long-standing quest to exploit the \textit{exponential representational capacity} of quantum systems for computational advantage. Feynman’s seminal vision in 1982~\cite{feynman1982simulating} first articulated the notion that a quantum system can naturally simulate another quantum system exponentially faster than any classical computer, inaugurating the field of quantum simulation. In parallel, the broader landscape of artificial intelligence evolved from logic-based systems (e.g., Newell and Simon’s \textit{Logic Theorist}, 1956~\cite{newell1956logic}) to statistical learning and deep neural networks (e.g., Hinton and Salakhutdinov’s deep belief networks, 2006~\cite{hinton2006reducing}). However, both paradigms---symbolic and statistical---share a fundamental limitation: they represent knowledge in discrete or real-valued spaces lacking the intrinsic notion of coherent superposition and interference.

Classical symbolic reasoning, as formalized in first-order logic and automated theorem proving~\cite{russell2010artificial}, relies on deterministic rule application over well-defined symbols. While precise and interpretable, these systems are brittle, unable to handle uncertainty or partial truth gracefully. The subsequent shift to statistical methods---notably Bayesian networks~\cite{pearl1988probabilistic}, graphical models~\cite{koller2009probabilistic}, and neural attention architectures~\cite{vaswani2017attention}---introduced probabilistic inference and large-scale pattern extraction. Yet, even the most advanced large language models (LLMs) remain fundamentally classical: their inference processes, though parallelized across vast dimensions, are linear in probabilities, not amplitudes. They cannot represent destructive interference or non-separable correlations between entities in the same expressive sense that quantum states can.

In contrast, quantum systems natively exhibit the essential ingredients of reasoning:
\begin{enumerate}
    \item \textbf{Superposition:} enabling simultaneous consideration of multiple hypotheses or possible worlds, akin to reasoning under uncertainty or counterfactual reasoning.
    \item \textbf{Entanglement:} encoding dependencies and implications between propositions, analogous to logical coupling or causal structure.
    \item \textbf{Interference:} allowing consistent hypotheses to constructively reinforce while contradictory ones destructively cancel, functioning as a built-in consistency filter.
    \item \textbf{Measurement:} producing a definite outcome only upon observation, mirroring the resolution of uncertainty when a decision or conclusion is reached.
\end{enumerate}

These properties suggest that quantum mechanics provides not only a computational substrate but also a natural formalism for reasoning. Historically, several researchers have speculated on this intersection. Finkelstein~\cite{finkelstein1969logic} and Birkhoff \& von Neumann~\cite{birkhoff1936logic} laid early theoretical foundations for \textit{quantum logic}, proposing a non-classical lattice of propositions where the distributive law of Boolean logic fails, reflecting the structure of quantum measurement outcomes. More recent quantum cognitive models~\cite{bruza2009quantum} and quantum probabilistic frameworks for decision making~\cite{busemeyer2012quantum} have used Hilbert space representations to capture human reasoning phenomena such as order effects and contextuality that violate classical probability.

However, despite these conceptual advances, few attempts have operationalized quantum principles for \textit{computational reasoning}. The majority of quantum algorithms still focus on optimization or pattern recognition rather than logical inference. The novelty of the present work lies in explicitly framing \textbf{reasoning as a quantum dynamical process}: inference becomes a unitary evolution of a superposition of truth assignments, driven by parameterized gates corresponding to logical rules and constraints. Logical consistency is enforced not through explicit rule evaluation, but via interference patterns emerging from the circuit’s structure.

In the proposed \textit{Quantum Circuit Reasoning Model (QCRM)}, propositions are encoded as qubits, rules as entangling unitaries, and contradictions as phase penalties. Measurement outcomes correspond to the probability amplitudes of consistent conclusions. This approach, unlike conventional neural-symbolic systems~\cite{garcez2019neural} or differentiable theorem provers~\cite{rocktaschel2017end}, replaces discrete rule application with continuous, reversible evolution, thereby maintaining the physical interpretability and auditability of the reasoning process.

By operating directly in a Hilbert space of propositions, QCRM provides a \textit{physically grounded alternative} to attention-based reasoning architectures. Whereas transformers perform reasoning through high-dimensional vector similarity and weighted averaging, quantum circuits compute reasoning through unitary transformations and global phase interactions. This shift from probabilistic inference to \emph{amplitude-based inference} introduces a fundamentally richer computational algebra: one that merges logic, geometry, and physics.

In summary, this work extends the scope of VQCs from numerical tasks to cognitive and logical domains, introducing the first end-to-end differentiable model where reasoning is expressed as quantum evolution. The motivation stems from both historical and conceptual insights: from Feynman’s call to simulate physics with physics, to Birkhoff and von Neumann’s vision of quantum logic, to modern neural reasoning’s search for coherence and interpretability. The resulting Quantum Circuit Reasoning Model represents a synthesis of these threads, offering a pathway toward physically interpretable, interference-driven reasoning machines.

\section{Foundations of Quantum Reasoning}

The formulation of reasoning within a quantum mechanical framework builds upon a lineage of ideas that merge logic, probability, and the physics of information. In classical artificial intelligence, a reasoning system is represented as a collection of atomic propositions and inference rules that determine how truth values propagate. However, such systems are inherently deterministic and Boolean in structure: every proposition is either true or false, and reasoning corresponds to the sequential manipulation of these discrete symbols. In contrast, quantum theory offers a fundamentally richer algebraic structure for representing information—one that naturally encompasses uncertainty, contextuality, and interference effects that have no classical analogue.

Historically, the connection between logic and quantum mechanics can be traced to the pioneering work of Birkhoff and von Neumann (1936) ~\cite{birkhoff1936logic}, who introduced the concept of \textit{quantum logic}. They proposed that the set of propositions about a quantum system forms not a Boolean algebra, as in classical logic, but an \textit{orthomodular lattice}, where the distributive law fails due to the contextual nature of measurement. This insight implied that the logical structure underlying quantum phenomena cannot be captured by classical reasoning frameworks. Later elaborations by Mackey~\cite{mackey1963mathematical} and Jauch~\cite{jauch1968foundations} established axiomatic treatments of quantum logic as an independent field of study.

Quantum reasoning thus emerges at the intersection of two intellectual traditions: (1) the study of logic as a formal system of inference, and (2) the mathematical foundations of quantum mechanics as developed by Dirac~\cite{dirac1958principles}, von Neumann~\cite{vonneumann1955mathematical}, and others. In the Hilbert-space formulation of quantum theory, physical states are represented as unit vectors in a complex vector space, and observables as Hermitian operators acting on that space. Probabilities are computed via the Born rule, which assigns the probability of an outcome to the squared magnitude of a complex amplitude. This inherently probabilistic yet linear structure has inspired numerous extensions of classical inference frameworks to the quantum domain.

Recent work in \textbf{quantum cognition} and \textbf{quantum probability}~\cite{busemeyer2012quantum,khrennikov2010ubiquitous,bruza2009quantum} has shown that many cognitive paradoxes—such as violations of the law of total probability, the conjunction fallacy, and contextual dependencies in human decision-making—can be modeled accurately using Hilbert space representations. These models leverage the superpositional nature of quantum states to represent uncertain or ambiguous mental states and use interference to model the constructive and destructive combination of evidence. In this sense, quantum probability provides a mathematically consistent extension of Bayesian reasoning into a non-commutative space.

Building upon this conceptual foundation, \textbf{Quantum Information Theory} introduced by Schumacher~\cite{schumacher1995quantum}, Nielsen and Chuang~\cite{nielsen2010quantum}, and others reframed information as a physical entity governed by quantum mechanics. The notion that logical propositions could themselves be encoded in qubits—two-level quantum systems—opened the possibility of reasoning as the manipulation of entangled quantum states. Entanglement, first recognized by Einstein, Podolsky, and Rosen~\cite{einstein1935can} and later formalized by Schrödinger~\cite{schrodinger1935discussion}, represents the quintessential non-classical correlation that allows distributed propositions to exhibit dependencies that defy separability. Within the context of reasoning, entanglement captures the idea of logical implication or causal dependency, where the truth of one proposition constrains or determines the truth of another.

Let a reasoning problem be expressed in terms of atomic propositions $\{p_i\}_{i=1}^N$, each mapped to a qubit representing its quantum state. The composite reasoning system is then described by the tensor product Hilbert space:
\[
\mathcal{H} = (\mathbb{C}^2)^{\otimes N},
\]
whose computational basis consists of the classical truth assignments $\{|x\rangle: x \in \{0,1\}^N\}$. A general reasoning state is a normalized superposition:
\begin{equation}
|\Psi\rangle = \sum_{x \in \{0,1\}^N} \psi_x |x\rangle, \quad \sum_x |\psi_x|^2 = 1,
\end{equation}
where $\psi_x \in \mathbb{C}$ are complex amplitudes. Each amplitude encodes not only the degree of belief in a particular truth assignment but also its phase, which governs how different hypotheses interfere when combined. 

In this representation:
\begin{itemize}
    \item The \textbf{superposition principle} allows simultaneous consideration of multiple hypotheses, embodying a form of parallel reasoning analogous to probabilistic reasoning but enriched by phase coherence.
    \item \textbf{Entanglement} enables the encoding of dependencies among propositions such that the truth value of one cannot be considered independently of others, representing logical coupling.
    \item \textbf{Unitary evolution} provides a reversible mechanism for inference, ensuring that information is preserved as reasoning proceeds through the quantum circuit.
    \item \textbf{Measurement} collapses the superposition into a definite truth assignment, corresponding to the resolution of uncertainty through decision or observation.
\end{itemize}

The formalism of reasoning in Hilbert space thus offers a synthesis of logical, probabilistic, and physical reasoning. It generalizes classical propositional logic into a \textit{quantum propositional calculus}, where inference corresponds to unitary transformations over complex amplitudes rather than discrete rule application. This approach extends the tradition of \textit{amplitude-based computation} (as in Feynman path integrals~\cite{feynman1948space}) into the domain of inference and symbolic reasoning. 

By conceptualizing reasoning as quantum state evolution, we can capture the dual aspects of logic (consistency and implication) and probability (uncertainty and evidence combination) within a unified mathematical structure. This represents a significant conceptual shift: from reasoning as deterministic symbol manipulation or statistical averaging to reasoning as structured interference in a high-dimensional Hilbert space. The Quantum Reasoning model developed in this work builds directly upon this foundation, using parameterized unitaries to evolve a reasoning state toward configurations that maximize logical consistency and coherence.

\subsection{Logical Propositions as Qubit States}
We map classical truth values to quantum basis states:
\[
|0\rangle \leftrightarrow \text{False}, \quad |1\rangle \leftrightarrow \text{True}.
\]
Uncertain beliefs are encoded as superpositions:
\[
|\psi_i\rangle = \alpha_i |0\rangle + \beta_i |1\rangle, \quad |\alpha_i|^2 + |\beta_i|^2 = 1.
\]
An initial belief state is then prepared as
\[
|\Psi_0\rangle = \bigotimes_{i=1}^{N} (\alpha_i |0\rangle + \beta_i |1\rangle).
\]
Here $\alpha_i$ and $\beta_i$ can be outputs of classical networks mapping prior evidence or features into rotation angles for state preparation.

\subsection{Logical Rules as Unitary Operations}
A logical rule such as $(A \land B) \Rightarrow C$ can be implemented as a multi-controlled rotation acting on qubit $C$, conditioned on $A$ and $B$:
\[
U_{(A,B)\rightarrow C}(\theta) = I - (1 - e^{-i\theta}) |11\rangle_{AB}\langle 11| \otimes \sigma_x^C,
\]
where $\sigma_x$ is the Pauli-X operator and $\theta$ controls the strength of the implication. This operation entangles the truth of $C$ with the conjunction of $A$ and $B$. Logical dependencies thus become quantum correlations.

\subsection{Consistency Constraints via Phase Penalties}
To penalize logically inconsistent configurations, we apply controlled phase shifts to those subspaces. For instance, if $A$ and $B$ cannot both be true, apply:
\[
U_{\text{penalty}}(\phi) = I - (1 - e^{i\phi}) |11\rangle_{AB}\langle 11|.
\]
Over successive layers, these phase rotations cause destructive interference among inconsistent amplitude paths, leaving self-consistent states with higher probability amplitudes.

\subsection{Inference as Quantum Evolution}
The full reasoning process is represented as a variational unitary evolution:
\begin{equation}
|\Psi(\bm{\theta})\rangle = U_L(\theta_L) \cdots U_2(\theta_2) U_1(\theta_1) |\Psi_0\rangle,
\end{equation}
where each $U_\ell(\theta_\ell)$ encodes entangling and phase-penalty gates derived from the rule structure. Measurement of target qubits yields the inferred truth probabilities:
\[
p_i = \langle \Psi(\bm{\theta}) | Z_i | \Psi(\bm{\theta}) \rangle,
\]
where $Z_i$ is the Pauli-Z operator on qubit $i$. These expectation values serve as the model's predictions for each queried proposition.

\section{Variational Optimization}
To train the model, we define a cost function based on target truth values $\{y_i\}$ for certain propositions:
\[
\mathcal{L}(\bm{\theta}) = \sum_i \text{BCE}(p_i(\bm{\theta}), y_i),
\]
where $\text{BCE}$ denotes binary cross-entropy loss. Classical optimization methods such as Adam or gradient descent are then used to update $\bm{\theta}$ via the parameter-shift rule:
\[
\frac{\partial \langle Z_i \rangle}{\partial \theta_k} = \frac{1}{2}[\langle Z_i \rangle_{\theta_k+\pi/2} - \langle Z_i \rangle_{\theta_k-\pi/2}].
\]
This results in a hybrid quantum-classical training loop capable of refining reasoning dynamics through data-driven optimization.

\section{Quantum Reasoning Layer (QRL) Architecture}

To operationalize the principles of quantum reasoning within a trainable computational framework, we introduce the \textbf{Quantum Reasoning Layer (QRL)}. The QRL is designed as a modular component that captures the dynamics of inference as the evolution of a multi-qubit quantum state through a parameterized sequence of unitaries. Each unitary sublayer encodes a distinct cognitive or logical function—rule propagation, consistency enforcement, and hypothesis exploration—implemented via entangling gates, phase shifts, and local rotations, respectively. The layer may be simulated efficiently on classical devices for small system sizes or executed natively on near-term quantum processors using standard gate sets.

\subsection{Mathematical Definition of the QRL}
Formally, a QRL acts on an input quantum state $|\Psi_0\rangle \in \mathcal{H} = (\mathbb{C}^2)^{\otimes N}$, where $N$ is the number of atomic propositions or qubits. The layer transforms this state through a parameterized unitary evolution:
\begin{equation}
|\Psi_{\text{out}}\rangle = U_{\text{QRL}}(\bm{\theta}, \bm{\phi}, \bm{\gamma}) |\Psi_0\rangle,
\end{equation}
where $\bm{\theta}$, $\bm{\phi}$, and $\bm{\gamma}$ denote the sets of trainable parameters corresponding to rule strength, constraint phase, and mixing rotations, respectively. The composite unitary is constructed as a product of sublayer unitaries:
\begin{equation}
U_{\text{QRL}}(\bm{\theta}, \bm{\phi}, \bm{\gamma}) = U_{\text{mix}}(\bm{\gamma}) \, U_{\text{penalty}}(\bm{\phi}) \, U_{\text{rule}}(\bm{\theta}),
\end{equation}
where each subcomponent acts on disjoint or overlapping subsets of qubits depending on the rule and constraint graph.

When $L$ such layers are stacked, we obtain a deep variational reasoning circuit:
\begin{equation}
U_{\text{QRL}}^{(L)} = \prod_{\ell=1}^{L} \left[ U_{\text{mix}}(\bm{\gamma}_\ell) \, U_{\text{penalty}}(\bm{\phi}_\ell) \, U_{\text{rule}}(\bm{\theta}_\ell) \right].
\end{equation}
This hierarchical structure is directly analogous to the depth of a neural network or the number of message-passing iterations in a Graph Neural Network (GNN), but extended into the complex-valued and reversible quantum domain.

\subsection{Entangling Rule Layer: Logical Propagation as Unitary Transformation}
The \textbf{Entangling Rule Layer} encodes logical dependencies among propositions by introducing multi-controlled rotations that couple qubits according to logical rules of the form:
\[
(p_i \land p_j) \Rightarrow p_k, \quad (p_i) \Rightarrow p_j, \quad (p_i \lor p_j) \Rightarrow p_k.
\]
Mathematically, a rule $(p_i, p_j) \Rightarrow p_k$ is represented by a unitary operator acting on three qubits:
\begin{equation}
U_{\text{rule}}^{(ijk)}(\theta) = \exp\left[-i \frac{\theta}{2} P_{ij} \otimes X_k \right],
\end{equation}
where $X_k$ is the Pauli-$X$ operator acting on target qubit $k$ (corresponding to logical negation or flipping of truth value) and $P_{ij} = |11\rangle_{ij}\langle 11|$ is a projector onto the subspace where both control qubits $i$ and $j$ are true. Expanding the exponential,
\[
U_{\text{rule}}^{(ijk)}(\theta) = I - (1 - \cos\frac{\theta}{2})P_{ij} + i\sin\frac{\theta}{2} P_{ij} X_k,
\]
shows that when $p_i = p_j = 1$, qubit $p_k$ undergoes a rotation about the $X$-axis by angle $\theta$. Thus, the amplitude of $p_k = 1$ is increased in proportion to the joint truth of $(p_i, p_j)$, analogously to a logical implication being “activated.” The parameter $\theta$ encodes the rule strength and is optimized during training.

For an entire reasoning graph $G = (V, E)$, with edges denoting implication relations, the global entangling operation is given by:
\begin{equation}
U_{\text{rule}}(\bm{\theta}) = \prod_{(i,j,k) \in \mathcal{R}} U_{\text{rule}}^{(ijk)}(\theta_{ijk}),
\end{equation}
where $\mathcal{R}$ is the set of all logical rules in the system.

\subsection{Phase Penalty Layer: Consistency Enforcement via Interference}
Logical consistency and physical feasibility constraints are enforced in the \textbf{Phase Penalty Layer} through the application of controlled phase shifts. The intuition is that inconsistent joint assignments (e.g., mutually exclusive propositions both being true) acquire a nonzero phase, leading to destructive interference when amplitudes are summed over all paths. For instance, to penalize configurations where $p_i$ and $p_j$ cannot both be true, we define:
\begin{equation}
U_{\text{penalty}}^{(ij)}(\phi) = \exp\left(i\phi |11\rangle_{ij}\langle 11|\right).
\end{equation}
Applying this unitary introduces a relative phase $e^{i\phi}$ to the $|11\rangle$ component of $(p_i, p_j)$. Over repeated applications within a circuit, amplitudes corresponding to logically inconsistent states destructively interfere, while consistent states constructively reinforce.

For a system with a set of pairwise and higher-order constraints $\mathcal{C}$, the composite penalty unitary is:
\begin{equation}
U_{\text{penalty}}(\bm{\phi}) = \prod_{(i,j) \in \mathcal{C}} U_{\text{penalty}}^{(ij)}(\phi_{ij}).
\end{equation}
In general, these constraints can encode exclusivity conditions (no two contradictory facts can be true), resource constraints (sum of probabilities must satisfy physical limits), or global logical invariants. In this manner, $U_{\text{penalty}}$ realizes a distributed form of constraint propagation through interference.

\subsection{Mixing Layer: Exploration and Amplitude Diffusion}
The \textbf{Mixing Layer} provides local degrees of freedom for exploration within the reasoning state space. Each qubit undergoes parameterized single-qubit rotations that allow the amplitudes to diffuse and prevent premature convergence to suboptimal local minima. The standard choice is a product of rotations about the $Y$ and $Z$ axes:
\begin{equation}
U_{\text{mix}}(\bm{\gamma}) = \prod_{i=1}^{N} R_z^{(i)}(\gamma_i^z) R_y^{(i)}(\gamma_i^y),
\end{equation}
where
\[
R_y(\theta) = \exp\left(-i\frac{\theta}{2} Y\right), \quad R_z(\theta) = \exp\left(-i\frac{\theta}{2} Z\right),
\]
and $Y$, $Z$ are Pauli matrices. The parameters $\gamma_i^y$ and $\gamma_i^z$ are trained to control the exploration-exploitation trade-off in the reasoning process: small values correspond to near-deterministic inference, while larger values maintain richer superpositions representing uncertainty and alternative hypotheses. This layer ensures the overall circuit remains expressive and capable of representing a diverse family of reasoning trajectories.

\subsection{Layer Composition and Depth}
The composition of entangling, phase, and mixing sublayers defines a single QRL block:
\[
U_{\text{QRL}}^{(1)} = U_{\text{mix}}(\bm{\gamma}_1) \, U_{\text{penalty}}(\bm{\phi}_1) \, U_{\text{rule}}(\bm{\theta}_1).
\]
For complex reasoning tasks requiring multi-hop inference or long-range logical propagation, multiple QRL blocks can be stacked:
\begin{equation}
U_{\text{QRL}}^{(L)} = \prod_{\ell=1}^{L} U_{\text{QRL}}^{(\ell)} = \prod_{\ell=1}^{L} \left[ U_{\text{mix}}(\bm{\gamma}_\ell) \, U_{\text{penalty}}(\bm{\phi}_\ell) \, U_{\text{rule}}(\bm{\theta}_\ell) \right].
\end{equation}
This iterative structure parallels the depth of classical reasoning models such as Transformer layers or GNN message-passing rounds, yet preserves unitarity and reversibility—key properties that allow interpretability and backtracking analysis. The circuit depth $L$ governs the expressive capacity of the reasoning process: shallow circuits correspond to single-step logical inference, while deeper ones can represent multi-hop deductions and global constraint satisfaction.

\subsection{Inference and Readout}
After the forward evolution through $U_{\text{QRL}}^{(L)}$, reasoning outcomes are obtained by measuring target qubits in the computational basis. The expected value of the Pauli-$Z$ observable for qubit $i$ is:
\begin{equation}
\langle Z_i \rangle = \langle \Psi_{\text{out}} | Z_i | \Psi_{\text{out}} \rangle.
\end{equation}
The probability that the proposition $p_i$ is true is therefore:
\begin{equation}
\hat{y}_i = \frac{1 - \langle Z_i \rangle}{2}.
\end{equation}
This mapping arises because $Z_i |0\rangle = |0\rangle$ and $Z_i |1\rangle = -|1\rangle$, implying that $\langle Z_i \rangle = +1$ for false (|0⟩) and $-1$ for true (|1⟩) states. Consequently, $\hat{y}_i$ represents a normalized probability estimate analogous to the sigmoid activation in neural networks but derived directly from a quantum expectation value.

For multi-qubit queries, joint probabilities can be computed using higher-order observables, e.g.,
\[
\langle Z_i Z_j \rangle = \langle \Psi_{\text{out}} | Z_i Z_j | \Psi_{\text{out}} \rangle,
\]
which measure the correlation between propositions $p_i$ and $p_j$. These correlations encode entanglement-induced dependencies and can be interpreted as learned relational beliefs between facts. The global reasoning output is then a vector of expectation values $\bm{\hat{y}} = (\hat{y}_1, \ldots, \hat{y}_N)$, which may be post-processed or compared against ground-truth labels using a classical loss function, such as binary cross-entropy.

\subsection{Interpretation and Computational Analogy}
The Quantum Reasoning Layer thus generalizes classical message-passing architectures to operate in complex Hilbert space. The \textit{Entangling Rule Layer} performs message passing via quantum correlations; the \textit{Phase Penalty Layer} enforces global coherence analogous to energy minimization in Hopfield networks; and the \textit{Mixing Layer} provides stochastic exploration akin to attention diffusion in Transformer architectures. Yet, unlike classical networks, the QRL preserves full reversibility and encodes logical structure through physically interpretable unitary operations. It provides a computational substrate where reasoning emerges as interference among amplitude flows, offering a path toward physically grounded, interpretable, and differentiable reasoning systems.

\section{Interpretation and Connection to Classical Reasoning}

The Quantum Circuit Reasoning Model (QCRM) provides an elegant physical analogue to classical reasoning systems, where logical inference emerges as a dynamical process governed by the laws of quantum mechanics rather than as a sequence of discrete rule applications. In this framework, reasoning corresponds to the controlled evolution of quantum amplitudes under unitary transformations that encode both dependencies among propositions and global constraints. Quantum interference acts as a natural mechanism for enforcing logical consistency: when different inference paths yield contradictory conclusions, their corresponding amplitudes destructively interfere, effectively canceling out inconsistent possibilities. Conversely, when multiple reasoning paths support the same conclusion, their amplitudes constructively reinforce each other, increasing the probability of that outcome upon measurement. This mechanism is analogous to how symbolic logic eliminates contradictory assignments during unification, or how attention mechanisms in deep learning selectively amplify semantically coherent representations.

A key feature distinguishing quantum reasoning from classical probabilistic reasoning lies in the complex nature of its amplitudes. In a Bayesian network or a neural model, reasoning is represented as real-valued probability propagation or weight-based averaging. By contrast, in the QCRM, amplitudes carry both magnitude and phase, allowing for the representation of interference phenomena that correspond to context-dependent or mutually exclusive reasoning outcomes. This provides a mechanism to model subtle dependencies and contradictions that classical logic or probability theory cannot naturally express, such as contextuality and non-commutativity of inference order---features reminiscent of human reasoning and linguistic interpretation~\cite{busemeyer2012quantum,khrennikov2010ubiquitous}.

Another critical property of the quantum reasoning framework is its \textbf{reversibility}. Because every operation in the QCRM is a unitary transformation, no information is ever lost during reasoning; rather, it is redistributed or phase-shifted within the Hilbert space. This stands in sharp contrast to most classical reasoning algorithms, which are inherently dissipative---intermediate computations are often discarded, and the reasoning process cannot be reversed to trace the precise origin of a conclusion. The unitarity of the QCRM enables complete auditability: each gate corresponds to a well-defined logical or relational operation whose contribution can be isolated, inspected, or even inverted. This property directly addresses one of the long-standing challenges in modern AI systems—\textit{interpretability}. Unlike deep neural networks, where the internal decision pathways are distributed and opaque, the quantum reasoning model allows for explicit decomposition of the reasoning trajectory in terms of physically meaningful transformations.

From a broader computational perspective, the QCRM offers a unification of symbolic and sub-symbolic reasoning paradigms within a single formalism. In symbolic systems, such as Prolog or resolution-based theorem provers, reasoning proceeds through discrete manipulation of logical clauses. In sub-symbolic systems like neural networks, reasoning is achieved through continuous transformations in high-dimensional vector spaces. Quantum reasoning lies at the intersection of these extremes: the quantum state space is continuous and differentiable, like in neural networks, but the transformations (gates) operate on discrete logical entities (qubits representing propositions). This hybrid nature allows the QCRM to maintain both the interpretability of symbolic reasoning and the flexibility of sub-symbolic computation, achieving what can be viewed as a \textit{quantum neuro-symbolic synthesis}.

The physical correspondences between quantum and classical reasoning concepts can be summarized as follows:
\begin{itemize}
    \item \textbf{Superposition $\leftrightarrow$ Hypothesis Branching:} In classical reasoning, a system often evaluates multiple potential hypotheses or possible worlds separately, such as in model checking or non-monotonic reasoning. In quantum reasoning, all possible hypotheses coexist simultaneously in a superposed state, with amplitudes encoding their relative plausibility. This allows for exponential parallelism in hypothesis evaluation within a single wavefunction.
    \item \textbf{Entanglement $\leftrightarrow$ Logical Coupling / Dependency:} In symbolic reasoning, dependencies between propositions are represented as logical rules or relational constraints. Entanglement captures this notion physically: once two qubits become entangled, the state of one cannot be specified independently of the other. This reflects the logical or causal coupling between dependent propositions, allowing the system to reason about implications, correlations, and joint truths in a fundamentally holistic manner.
    \item \textbf{Interference $\leftrightarrow$ Consistency Filtering:} In classical inference, consistency is maintained by pruning contradictory models through constraint satisfaction or probabilistic normalization. In the quantum model, destructive interference plays this role: inconsistent amplitude paths cancel out through phase opposition, leaving only self-consistent states with nonzero amplitude. Thus, consistency emerges dynamically from wave interference rather than algorithmic filtering.
    \item \textbf{Measurement $\leftrightarrow$ Decision / Belief Collapse:} The act of measurement in quantum mechanics collapses a superposed state into a single observed outcome. Analogously, in reasoning, measurement corresponds to committing to a specific conclusion or belief given the superposed hypotheses. This process is inherently probabilistic—the frequency of outcomes reflects the likelihood of each consistent explanation, mirroring Bayesian decision-making but arising from amplitude statistics rather than real-valued probabilities.
\end{itemize}

This mapping highlights how quantum mechanics provides not merely a computational advantage but a profound conceptual framework for reasoning. It reframes inference as a process of \textit{coherent evolution} rather than sequential rule application. Logical dependencies are encoded in entangling gates; contradictions are resolved through interference; and conclusions emerge as measurement outcomes weighted by the constructive combination of consistent reasoning paths. The QCRM thus represents an interpretable, reversible, and physically grounded form of inference that transcends both classical symbolic reasoning and probabilistic neural computation, suggesting a new paradigm for explainable and self-consistent machine reasoning systems.

\section{Simulation on Classical Hardware}

While the long-term vision of the Quantum Circuit Reasoning Model (QCRM) is to operate natively on quantum hardware, current technological limitations—such as limited qubit coherence times, gate fidelity, and qubit count—necessitate practical exploration through \textbf{classical simulation}. Fortunately, the mathematical formalism of quantum mechanics is entirely linear and differentiable, allowing us to simulate the behavior of quantum systems on classical computers using modern tensor libraries such as \texttt{PyTorch}, \texttt{TensorFlow}, or \texttt{JAX}. These frameworks natively support complex-valued tensors and automatic differentiation, making them particularly suitable for implementing hybrid quantum-classical learning models. This approach enables the development, debugging, and training of QCRM architectures before their deployment on real quantum devices.

\subsection{Dimensionality and Practical Constraints}

The principal challenge in simulating quantum systems classically arises from the exponential scaling of the Hilbert space. For $N$ qubits, the state vector $|\Psi\rangle$ resides in a complex vector space of dimension $2^N$. This implies that storing and manipulating the full wavefunction requires memory and computational resources that scale as $\mathcal{O}(2^N)$. Despite this exponential growth, practical reasoning experiments can often be conducted with small systems ($N \leq 12$ qubits), which are sufficient to capture nontrivial logical dependencies and interference effects while remaining computationally tractable on modern GPUs. For instance, a 12-qubit state vector requires $2^{12} = 4096$ complex amplitudes, which can be efficiently handled on a single workstation.

To mitigate the exponential cost for larger systems, tensor factorization techniques such as matrix product states (MPS), tensor networks, or low-rank approximations can be employed. These techniques exploit the structured entanglement of reasoning circuits, representing the global state as a composition of locally entangled tensors. While not exact, such approximations preserve most of the relevant correlations for reasoning tasks with limited depth or locality of rules, making them practical for large-scale simulations.

\subsection{Differentiable Simulation Pipeline}

A classical simulation of the QCRM involves four key computational stages, each of which maps directly to a step in the quantum reasoning process. These stages can be implemented efficiently using GPU-accelerated tensor operations.

\paragraph{1. State Vector Initialization from Classical Features.}  
The reasoning process begins by encoding the initial beliefs or feature-derived priors into a quantum state. Given a set of classical inputs $\mathbf{x} = (x_1, x_2, \dots, x_N)$ representing prior knowledge about propositions, each qubit $p_i$ is initialized to a superposition:
\[
|\psi_i\rangle = \cos\left(\frac{\pi x_i}{2}\right)|0\rangle + \sin\left(\frac{\pi x_i}{2}\right)|1\rangle,
\]
which maps the normalized feature $x_i \in [0,1]$ to a point on the Bloch sphere. The full reasoning state is constructed as a tensor product:
\[
|\Psi_0\rangle = \bigotimes_{i=1}^{N} |\psi_i\rangle.
\]
This preparation step effectively embeds classical data into the quantum Hilbert space, providing a differentiable link between neural encoders and quantum reasoning modules.

\paragraph{2. Application of Parameterized Gates Using Kronecker Embeddings.}  
Each reasoning layer consists of unitary gates—entangling, phase, and mixing operations—applied to subsets of qubits. On classical hardware, these gates are represented as complex-valued matrices of size $2^N \times 2^N$. The global unitary $U_{\text{QRL}}$ can be efficiently constructed through Kronecker (tensor) products and sparse operator embeddings. For instance, a single-qubit rotation $R_y^{(i)}(\theta)$ acting on qubit $i$ is embedded into the full Hilbert space via:
\[
U^{(i)} = I_1 \otimes I_2 \otimes \cdots \otimes R_y(\theta_i) \otimes \cdots \otimes I_N,
\]
where each $I_j$ is a $2 \times 2$ identity matrix. Multi-controlled gates, such as those encoding logical implications, can be implemented using block-diagonal matrices or via conditional masking operations in tensor algebra. This modular representation allows the efficient composition of rule-based and constraint-based unitaries, forming the backbone of the simulated QRL circuit.

\paragraph{3. Computation of Observables via Complex Inner Products.}  
After the state has evolved under the sequence of unitaries, reasoning outcomes are inferred by computing the expectation values of relevant observables. For each qubit $i$, the expectation of the Pauli-$Z$ operator provides the estimated truth probability of the associated proposition:
\[
\langle Z_i \rangle = \langle \Psi_{\text{out}} | Z_i | \Psi_{\text{out}} \rangle,
\quad \hat{y}_i = \frac{1 - \langle Z_i \rangle}{2}.
\]
Higher-order correlations between propositions (e.g., co-dependencies or joint truth probabilities) can be captured through pairwise or tensor-product observables such as $\langle Z_i Z_j \rangle$ or $\langle Z_i Z_j Z_k \rangle$. These inner products are computed directly using tensor contraction operations, which are fully differentiable in frameworks like PyTorch or JAX.

\paragraph{4. Backpropagation Through Unitaries Using Automatic Differentiation or the Parameter-Shift Rule.}  
The differentiability of the quantum reasoning model is essential for training its parameters. Modern automatic differentiation engines support complex gradients, enabling end-to-end backpropagation through the entire circuit. Alternatively, the \textit{parameter-shift rule}~\cite{schuld2019evaluating} provides an exact gradient estimation method for quantum gates:
\[
\frac{\partial \langle Z_i \rangle}{\partial \theta_k} = \frac{1}{2}\left[\langle Z_i \rangle_{\theta_k + \pi/2} - \langle Z_i \rangle_{\theta_k - \pi/2}\right].
\]
This rule can be directly implemented in simulation, ensuring exact gradient computation without numerical instability. The resulting gradient signals are then used by classical optimizers (e.g., Adam, RMSProp) to update circuit parameters $\bm{\theta}, \bm{\phi}, \bm{\gamma}$ during training.

\subsection{Hybrid Quantum-Classical Training}

The differentiable simulation pipeline enables hybrid integration between quantum reasoning modules and classical neural networks. For instance, a deep neural encoder can map raw sensory or textual inputs to a latent vector $\mathbf{z}$, which is then used to parameterize the initial quantum state $|\Psi_0(\mathbf{z})\rangle$. The output of the quantum circuit, given by the expectation values $\bm{\hat{y}}$, can then be fed into a downstream classifier or loss function. The entire system is trained end-to-end by propagating gradients through both the classical and quantum components. This hybrid architecture leverages the representational power of quantum reasoning while maintaining the scalability and hardware accessibility of classical computation.

\subsection{Advantages of Classical Simulation}

Simulating QCRM on classical hardware provides several immediate advantages. It enables rapid prototyping of circuit designs, visualization of reasoning dynamics through amplitude and phase trajectories, and empirical testing of rule-encoding strategies. Moreover, because classical simulation supports exact wavefunction inspection, researchers can analyze the intermediate reasoning states—something that is generally impossible on physical quantum hardware due to measurement collapse. This capability allows detailed debugging, interpretability analysis, and benchmarking of reasoning consistency across circuit depths and datasets.

In summary, classical simulation serves as both a developmental and conceptual bridge toward full quantum deployment. It allows researchers to explore the computational and cognitive implications of quantum reasoning models using readily available hardware and modern deep learning libraries. As quantum processors mature, these simulated architectures can be transferred seamlessly to real quantum devices, enabling a future where logical inference, learning, and decision-making are unified within the physics of computation.

\section{Applications and Future Work}

The Quantum Circuit Reasoning Model (QCRM) represents a foundational shift in how reasoning can be conceived computationally. By expressing logical inference as unitary evolution in a quantum Hilbert space, the framework provides both a physical and mathematical substrate for coherent, explainable, and reversible reasoning. Unlike large language models (LLMs) based on attention mechanisms—which perform reasoning implicitly through high-dimensional token associations—QCRM performs reasoning explicitly through structured transformations of amplitudes representing propositions. This section elaborates on the key application domains where this model can provide transformative advantages, as well as the future research directions and limitations associated with scaling quantum reasoning architectures relative to current attention-based models.

\subsection{Applications}

\paragraph{1. Chemical Inference.}  
One of the most promising near-term applications of QCRM lies in \textbf{chemical and materials inference}, particularly for problems where reasoning about molecular interactions or reaction outcomes involves uncertainty and combinatorial complexity. For instance, predicting whether a pair of compounds forms a eutectic mixture depends on nontrivial interactions between hydrogen bond donors (HBDs) and acceptors (HBAs). Traditional machine learning models, including attention-based graph transformers and ChemBERTa-style language models for SMILES strings, treat this as a regression or classification problem over static embeddings. In contrast, a quantum reasoning model can encode the donor–acceptor relationships as entangled qubits, where each logical rule (e.g., "if both compounds exhibit complementary polarity, eutectic formation is likely") corresponds to an entangling gate. Interference among different molecular configurations automatically filters out physically inconsistent combinations, allowing for more interpretable predictions that align with chemical intuition. This approach bridges symbolic domain knowledge with statistical learning, offering a more transparent reasoning mechanism for chemical design and discovery.

\paragraph{2. Clinical Decision Support.}  
In the biomedical domain, reasoning under uncertainty is a fundamental challenge. Clinical decision-making often involves integrating heterogeneous sources of evidence—symptoms, lab results, imaging findings—into diagnostic hypotheses. Current clinical LLMs (e.g., Med-PaLM, BioGPT, or GPT-4-based diagnostic systems) rely heavily on language context and pattern completion but struggle with causal coherence and probabilistic consistency. The QCRM, by contrast, can represent causal and diagnostic graphs as quantum reasoning circuits. Propositions such as “symptom A and B imply condition C” can be encoded as entangling rules, while contradictory findings (e.g., “C and D cannot co-occur”) can be handled through phase penalties that enforce logical consistency via destructive interference. Measurement of the resulting amplitudes yields probabilistic diagnoses that are both interpretable and dynamically updated as new evidence becomes available. This could form the foundation of a new class of quantum-assisted \textit{clinical decision support systems} that reason over uncertainty in a physically interpretable manner, combining the strengths of Bayesian inference and causal modeling with the computational richness of quantum amplitude evolution.

\paragraph{3. Symbolic Reasoning and Cognitive Benchmarks.}  
Another compelling application area is \textbf{symbolic reasoning and entailment simulation}. Datasets such as ProofWriter, EntailmentBank, and CLUTRR are benchmarks designed to evaluate multi-hop reasoning, compositional generalization, and logical entailment capabilities of machine learning systems. While large transformer-based LLMs can solve many of these tasks through memorization and heuristic pattern completion, they often fail to generalize when the reasoning structure changes or when trained rules must be recombined in novel ways. The QCRM framework offers a new avenue: reasoning emerges naturally from unitary dynamics where entangled amplitudes represent interdependent propositions, and interference ensures global consistency. Logical entailments correspond to amplitude amplification in consistent subspaces, providing a physically interpretable analogue to proof search. In this sense, QCRM could serve as a quantum counterpart to neural-symbolic reasoning systems, with the additional benefit of explicit reversibility and explainability.

\section{Architectural Diagram: Quantum Circuit Reasoning Model (QCRM)}

In this section we provide a TikZ-based architectural outline of the Quantum Circuit Reasoning Model (QCRM). The goal of the diagram is to (1) depict the full signal flow from classical inputs to quantum state preparation, variational reasoning, and readout, and (2) annotate each transformation with its mathematical role. The diagram is modularized into four conceptual stages:

\begin{enumerate}
    \item \textbf{Classical Feature Encoding:} maps observed evidence or prior beliefs into continuous parameters.
    \item \textbf{Quantum State Preparation:} initializes a multi-qubit superposition $|\Psi_0\rangle$ from these parameters.
    \item \textbf{Quantum Reasoning Layer Stack:} performs iterative rule propagation, constraint enforcement, and amplitude mixing using parameterized unitaries.
    \item \textbf{Measurement and Classical Output:} extracts predicted truth values and correlations between propositions via expectation values of observables.
\end{enumerate}

\begin{figure}[ht!]
\centering
\resizebox{\textwidth}{!}{%
\begin{tikzpicture}[
    node distance=2.5cm,
    every node/.style={font=\small},
    module/.style={draw, rounded corners, thick, align=center,
                    minimum width=4.2cm, minimum height=1.3cm, fill=gray!5},
    quantum/.style={draw, rounded corners, thick, align=center,
                    minimum width=4.2cm, minimum height=1.3cm, fill=blue!5},
    flow/.style={-Stealth, thick},
]

\node[module] (input) {
    \textbf{Classical Input Features} \\
    Evidence / Observations $\mathbf{x} = (x_1,\ldots,x_N)$
};

\node[module, right=of input] (encoder) {
    \textbf{Classical Encoder / Prior Mapper} \\
    $\alpha_i, \beta_i \gets f_\text{enc}(x_i)$ \\
    $\theta_i^y = \pi x_i,\ \theta_i^z = g(x_i)$
};

\node[quantum, right=of encoder] (prep) {
    \textbf{Quantum State Preparation} \\
    $|\psi_i\rangle = R_z(\theta_i^z)R_y(\theta_i^y)|0\rangle$ \\
    $|\Psi_0\rangle = \bigotimes_i |\psi_i\rangle$
};

\node[quantum, right=3.8cm of prep] (qrl1) {
    \textbf{QRL Block 1} \\
    $U_{\text{QRL}}^{(1)} = U_{\text{mix}}U_{\text{pen}}U_{\text{rule}}$
};

\node[quantum, below=1.8cm of qrl1] (qrl2) {
    \textbf{QRL Block 2} \\
    $U_{\text{QRL}}^{(2)} = U_{\text{mix}}U_{\text{pen}}U_{\text{rule}}$
};

\node[quantum, below=1.8cm of qrl2] (qrll) {
    \textbf{QRL Block $L$} \\
    $U_{\text{QRL}}^{(L)} = U_{\text{mix}}U_{\text{pen}}U_{\text{rule}}$
};

\draw[decorate,decoration={brace,amplitude=6pt,mirror}]
(qrll.south west) -- (qrl1.north west)
node[midway,xshift=-3.4cm,align=center,rotate=90,text width=3cm]
{\textbf{Quantum Reasoning}\\\textbf{Layer Stack}\\(Entanglement + Interference)};

\node[module, right=4.0cm of qrl2] (measure) {
    \textbf{Measurement / Readout} \\
    $\langle Z_i \rangle = \langle \Psi_{\text{out}}|Z_i|\Psi_{\text{out}}\rangle$ \\
    $\hat{y}_i = \frac{1-\langle Z_i\rangle}{2}$
};

\node[module, right=of measure] (output) {
    \textbf{Classical Output / Loss} \\
    Predictions $\hat{y}_i$ \\
    Loss $\mathcal{L}(\theta,\phi,\gamma)$
};

\draw[flow] (input) -- (encoder);
\draw[flow] (encoder) -- (prep);
\draw[flow] (prep) -- node[above]{Initial state $|\Psi_0\rangle$} (qrl1);
\draw[flow] (qrl1.south) -- (qrl2.north);
\draw[flow] (qrl2.south) -- (qrll.north);
\draw[flow] (qrl2.east) -- node[above]{Final $|\Psi_{\text{out}}\rangle$} (measure.west);
\draw[flow] (measure) -- (output);

\end{tikzpicture}
}
\caption{Architectural signal flow of the Quantum Circuit Reasoning Model (QCRM). Classical data is encoded into qubit rotations, evolved through a stack of Quantum Reasoning Layers (QRLs), and measured to yield interpretable logical outputs.}
\label{fig:arch}
\end{figure}
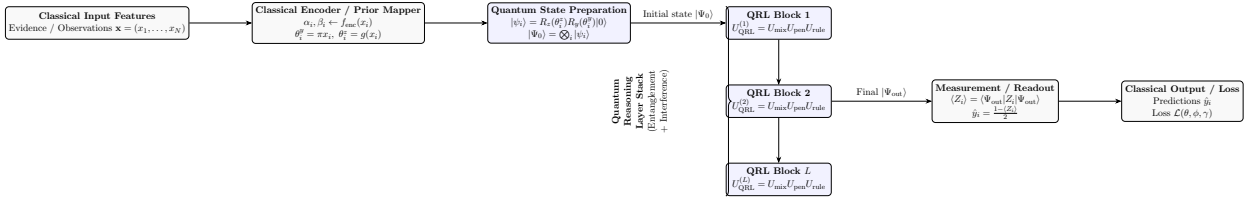

Below we present an architectural diagram that encodes this pipeline in Fig.~\ref{fig:arch}. After the code, we include explanatory text describing each component and its mathematical semantics.

\subsection{Narrative Walkthrough of the Architecture}

\paragraph{1. Classical Input $\rightarrow$ Encoder.}
The pipeline begins with structured or unstructured evidence $\mathbf{x} = (x_1,\ldots,x_N)$, which may include symbolic facts, graph features, clinical observations, molecular descriptors, or contextual priors. These inputs are passed through a classical encoder $f_\text{enc}$ (often a shallow MLP, a GNN, or a rule-derived mapping) that outputs continuous parameters for each proposition/qubit. For each proposition $p_i$, this encoder yields rotation angles $(\theta_i^y, \theta_i^z)$ or equivalently amplitudes $(\alpha_i,\beta_i)$ specifying how much prior belief mass we assign to $p_i = 0$ vs. $p_i = 1$. Formally,
\[
(\theta_i^y, \theta_i^z) = f_\text{enc}(x_i), \qquad
|\psi_i\rangle = R_z(\theta_i^z) R_y(\theta_i^y) |0\rangle
= \alpha_i |0\rangle + \beta_i |1\rangle.
\]
This defines a differentiable interface between classical perception and quantum reasoning.

\paragraph{2. State Preparation.}
All single-qubit states $|\psi_i\rangle$ are combined into a global initial reasoning state $|\Psi_0\rangle$ via tensor product:
\[
|\Psi_0\rangle = \bigotimes_{i=1}^{N} |\psi_i\rangle,
\quad
|\psi_i\rangle = \alpha_i |0\rangle + \beta_i |1\rangle,
\quad
|\alpha_i|^2 + |\beta_i|^2 = 1.
\]
This state encodes, in superposition, all $2^N$ joint assignments of truth values to the $N$ propositions $\{p_i\}$. The amplitude of each assignment reflects both local priors (from individual $p_i$) and correlations implicitly induced by shared encoder structure. At this point, the system has \emph{not} yet enforced any global logical structure; it is a preparation of belief amplitudes.

\paragraph{3. Quantum Reasoning Layer Stack.}
The heart of the model is the repeated application of $L$ Quantum Reasoning Layers. Each QRL block applies a composite unitary
\[
U_{\text{QRL}}^{(\ell)} = U_{\text{mix}}^{(\ell)} \, U_{\text{pen}}^{(\ell)} \, U_{\text{rule}}^{(\ell)},
\]
so that after $L$ blocks, the output state is
\[
|\Psi_{\text{out}}\rangle
= \left( \prod_{\ell=1}^{L} U_{\text{QRL}}^{(\ell)} \right) |\Psi_0\rangle.
\]

Within each block:
\begin{itemize}
    \item $U_{\text{rule}}$ (\textbf{Entangling Rule Layer}) applies multi-controlled unitaries that realize implications such as $(p_i \land p_j) \Rightarrow p_k$. A typical operation is
    \[
    U_{\text{rule}}^{(ijk)}(\theta) = \exp\!\left[-i \frac{\theta}{2} \left(|11\rangle_{ij}\langle 11| \otimes X_k \right)\right],
    \]
    which increases the amplitude of $p_k = 1$ \emph{conditioned} on $p_i = p_j = 1$. This is how relational structure and logical coupling are injected into the joint amplitude distribution.
    \item $U_{\text{pen}}$ (\textbf{Phase Penalty Layer}) applies controlled phase shifts to states violating constraints. For example, for an inconsistent pair $(p_a, p_b)$ that cannot both be true:
    \[
    U_{\text{pen}}^{(ab)}(\phi) = \exp\!\left(i \phi |11\rangle_{ab}\langle 11|\right).
    \]
    Over repeated layers, these phase kicks induce destructive interference among logically inconsistent global assignments, suppressing them in $|\Psi_{\text{out}}\rangle$.
    \item $U_{\text{mix}}$ (\textbf{Mixing Layer}) applies local rotations such as
    \[
    R_y(\gamma_i^y) = e^{-i \frac{\gamma_i^y}{2} Y}, \quad
    R_z(\gamma_i^z) = e^{-i \frac{\gamma_i^z}{2} Z},
    \]
    to each qubit $p_i$. This prevents premature collapse into overly sharp hypotheses and enables exploration of alternative consistent assignments. It is analogous to temperature or diffusion in probabilistic inference.
\end{itemize}

Conceptually, $U_{\text{rule}}$ propagates implications (``if this, then that''), $U_{\text{pen}}$ prunes contradictions through interference, and $U_{\text{mix}}$ maintains uncertainty and encourages hypothesis diversity. Stacking these blocks performs iterative multi-hop inference, similar to message passing in a GNN or depth in a Transformer, but now with full unitarity and reversibility.

\paragraph{4. Measurement and Observable Extraction.}
After propagating through all QRL blocks, we obtain $|\Psi_{\text{out}}\rangle$. The model does not immediately ``collapse'' in simulation; instead, we compute expectation values of observables. For each proposition qubit $p_i$, we read out its predicted truth probability using
\[
\langle Z_i \rangle = \langle \Psi_{\text{out}} | Z_i | \Psi_{\text{out}} \rangle,
\qquad
\hat{y}_i = \frac{1 - \langle Z_i \rangle}{2}.
\]
Because $Z|0\rangle = |0\rangle$ and $Z|1\rangle = -|1\rangle$, we have $\langle Z_i \rangle = +1$ if $p_i$ is certainly false and $-1$ if $p_i$ is certainly true; $\hat{y}_i$ linearly rescales this to $[0,1]$. For reasoning about relationships between propositions $p_i$ and $p_j$, we can compute higher-order correlators such as
\[
\langle Z_i Z_j \rangle = \langle \Psi_{\text{out}} | Z_i Z_j | \Psi_{\text{out}} \rangle,
\]
which capture entanglement-induced dependencies. These correlations serve as an interpretable diagnostic of whether two facts are expected to co-occur, be mutually exclusive, or be causally linked in the inferred explanation.

\paragraph{5. Classical Loss and Differentiable Training.}
Finally, the predicted beliefs $\{\hat{y}_i\}$ (and potentially pairwise expectations or causal indicators) are compared to supervision signals or desired reasoning outcomes. We define a classical loss $\mathcal{L}$ such as binary cross-entropy:
\[
\mathcal{L}(\theta,\phi,\gamma)
= \sum_i \text{BCE}\big(\hat{y}_i, y_i^\text{target}\big),
\]
and update the parameters of $U_{\text{rule}}$, $U_{\text{pen}}$, and $U_{\text{mix}}$ using gradients. Gradients can be obtained either via automatic differentiation in a classical simulator (treating the complex amplitudes as differentiable tensors) or via parameter-shift rules of the form
\[
\frac{\partial \hat{y}_i}{\partial \theta_k}
= \frac{1}{2}\left[
\hat{y}_i(\theta_k + \frac{\pi}{2}) -
\hat{y}_i(\theta_k - \frac{\pi}{2})
\right].
\]
This closes the learning loop: the circuit does not merely \emph{apply} logic, it \emph{learns} how strongly to apply each rule, how harshly to penalize contradictions, and how much exploratory mixing to allow.

\subsection{Summary of Signal Semantics}

\begin{itemize}
    \item \textbf{Classical input $\rightarrow$ angles.} Raw factual/observational data is turned into qubit initialization parameters, representing prior belief amplitudes.
    \item \textbf{Angles $\rightarrow |\Psi_0\rangle$.} These parameters instantiate a full joint superposition of all candidate worlds.
    \item \textbf{$|\Psi_0\rangle \rightarrow |\Psi_{\text{out}}\rangle$.} A stack of QRL blocks applies learnable logical propagation, global consistency enforcement (via interference), and controlled uncertainty diffusion.
    \item \textbf{$|\Psi_{\text{out}}\rangle \rightarrow \hat{y}$.} Expectation values of observables map amplitudes back to classical predicted truth values and relational structure.
    \item \textbf{$\hat{y} \rightarrow \mathcal{L} \rightarrow \nabla$.} Supervised targets drive gradient updates of all QRL parameters, giving a fully differentiable, end-to-end trainable reasoning engine.
\end{itemize}

This TikZ architecture formalizes QCRM not just as an abstract idea but as an executable pipeline: classical evidence in, quantum-evolved multi-hypothesis reasoning in the middle, and interpretable probabilistic beliefs out.

\subsection{Future Work}

\paragraph{1. Incorporating Noise Models for Probabilistic Reasoning.}  
In real-world applications, information is often noisy or incomplete. Future extensions of QCRM should incorporate stochastic noise channels—such as depolarizing, amplitude-damping, or phase-flip models—to represent uncertainty in observations or rule reliability. These noise models transform the reasoning process from deterministic unitary evolution to a mixture of quantum operations described by completely positive trace-preserving (CPTP) maps. The resulting “open quantum reasoning” framework could more accurately reflect the probabilistic nature of human reasoning and decision-making, drawing analogies to cognitive uncertainty and bounded rationality.

\paragraph{2. Defining Rule Templates as Parameterized Hamiltonians.}  
Another promising direction is to represent logical rules not merely as discrete gates but as continuous-time evolutions governed by parameterized Hamiltonians. In this view, reasoning corresponds to minimizing an energy functional analogous to the consistency energy of a logic network. A Hamiltonian term $H_{(i,j,k)}$ could encode the constraint that “if $p_i$ and $p_j$ are true, $p_k$ should also be true,” and inference would correspond to evolving the system under the combined Hamiltonian $H = \sum H_{(i,j,k)}$ until it reaches a low-energy, self-consistent state. This connects the QCRM framework to energy-based models in machine learning, such as Hopfield networks or Boltzmann machines, while preserving the physical interpretability and reversibility of quantum evolution.

\paragraph{3. Compositional Generalization and Circuit Reuse.}  
A major limitation of current reasoning models—including large attention-based transformers—is their difficulty with compositional generalization: reusing learned reasoning structures in novel combinations. QCRM provides a principled mechanism for compositionality by defining reusable circuit templates corresponding to reasoning motifs (e.g., conjunction, implication, exclusion). These circuit fragments can be reassembled dynamically to handle new logical graphs without retraining the entire model, much like modular quantum subroutines. Exploring how these templates can be automatically discovered, stored, and recombined is a key area for future research and may yield models capable of one-shot or few-shot reasoning transfer across domains.

\paragraph{4. Integration with Neural Embeddings for Perception-to-Reasoning Transfer.}  
In hybrid quantum-classical architectures, perceptual modules—such as vision transformers or language encoders—can provide embeddings that initialize the quantum reasoning state. The challenge lies in designing differentiable interfaces between continuous classical embeddings and discrete quantum propositions. Future work should investigate how latent representations from neural networks can parameterize qubit states and rule strengths, enabling end-to-end learning across perception and reasoning pipelines. This hybridization could enable systems that perceive complex data (e.g., clinical imaging or molecular graphs) and reason over them quantum-mechanically, bridging the gap between sub-symbolic and symbolic processing in a physically unified framework.

\subsection{Comparative Discussion: Advantages and Limitations}

Compared to attention-based large language models, QCRM offers several conceptual and practical advantages. First, it introduces \textbf{explicit consistency enforcement} through quantum interference, where contradictions are resolved physically rather than statistically. Second, it is inherently \textbf{interpretable and reversible}: each gate corresponds to a distinct logical operation, and the full inference trajectory can be reconstructed, providing a degree of transparency that attention weights or hidden embeddings lack. Third, the model possesses \textbf{built-in parallelism} through superposition—allowing simultaneous reasoning over exponentially many hypotheses within a single state vector, something classical architectures can only approximate through massive parameter scaling.

However, there are also important limitations. The exponential scaling of Hilbert space restricts classical simulation to small systems, making direct application to large-scale reasoning tasks computationally demanding. While future quantum hardware may alleviate this, near-term realizations will remain hybrid and limited in qubit count. Moreover, QCRM’s reasoning is fundamentally linear and unitary; without explicit noise or decoherence modeling, it lacks the stochastic adaptability of probabilistic attention models that can handle incomplete or contradictory data gracefully. Finally, integrating language-based symbolic abstraction into quantum circuits remains an open challenge: while attention-based LLMs excel at representing rich linguistic context, QCRM must develop mechanisms to translate semantic embeddings into qubit-level logical structures.

In summary, the Quantum Circuit Reasoning Model presents a radically new paradigm for reasoning—combining the interpretability and structure of symbolic logic with the expressivity and parallelism of quantum mechanics. While still in its infancy relative to the vast scale of transformer-based reasoning systems, QCRM offers a path toward physically grounded, transparent, and consistent reasoning architectures. As quantum computing technologies mature, this framework may evolve into a practical alternative or complement to large language models, redefining how machines reason, infer, and understand in the quantum age.

\section{Conclusion}

The development of  Quantum Circuit Reasoning Models (QCRM) marks a significant conceptual leap in the evolution of artificial intelligence architectures. Unlike traditional symbolic systems that rely on discrete rule application, or attention-based large language models that approximate reasoning through statistical correlations, QCRM provides a  physics-grounded  framework in which reasoning emerges as the unitary evolution of quantum amplitudes. By encoding logical relations as entangling operations and enforcing global consistency through quantum interference, the model transforms inference into a process of coherent state evolution rather than heuristic pattern completion. This reconceptualization situates reasoning not as a symbolic abstraction or probabilistic estimate, but as a physical computation governed by the foundational principles of quantum mechanics.

The framework offers several key advantages that distinguish it from current reasoning paradigms. First, it combines the  interpretability and structure  of symbolic logic with the  differentiability and trainability  of deep learning. Each unitary gate corresponds to an explicit logical or relational transformation, enabling full traceability and transparency throughout the reasoning process. Second, QCRM leverages the inherent  parallelism of quantum superposition, allowing simultaneous exploration of exponentially many hypotheses within a unified Hilbert space. Third,  interference serves as a built-in mechanism for consistency enforcement , automatically suppressing contradictory reasoning paths while amplifying coherent ones—an ability that current attention-based LLMs only approximate through learned soft weighting. Finally, the reversible and lossless nature of unitary evolution preserves all intermediate information, supporting explainable backtracking and verification—capabilities essential for trustworthy reasoning systems in high-stakes domains such as medicine, chemistry, and scientific discovery.

From an applied perspective, QCRM has the potential to bridge the gap between statistical learning and formal reasoning. In chemical inference, it provides a natural way to represent molecular interactions and rule-based composition through entanglement; in clinical decision support, it offers a mechanism for integrating uncertain and causal evidence within a coherent probabilistic logic; and in symbolic reasoning benchmarks, it demonstrates the possibility of performing compositional, multi-hop inference using quantum amplitude logic rather than attention-based heuristics. As such, the model unites the precision of symbolic AI, the flexibility of neural networks, and the physical realism of quantum mechanics into a single coherent framework.

Looking forward, the true impact of QCRM will unfold as quantum hardware matures and hybrid quantum-classical learning pipelines become more accessible. Near-term simulations on classical hardware have already shown that small-scale quantum reasoning circuits can model logical dependencies and consistency relationships in ways that are interpretable, differentiable, and compatible with neural encoders. As qubit counts and coherence times increase, these models can be scaled to represent larger reasoning graphs, more complex causal dependencies, and domain-specific rule systems. In the long term, QCRM may serve as the foundation for a new generation of  physically interpretable AI systems, machines that do not merely emulate reasoning statistically, but embody it as a natural process of physical inference within the laws of quantum mechanics.

In summary, Quantum Circuit Reasoning Models demonstrate that the frontiers of reasoning and computation can converge within a single mathematical and physical paradigm. By viewing inference as a coherent, reversible, and interference-driven process, QCRM establishes a bridge between logic, learning, and physics. It stands as a compelling step toward the creation of reasoning systems that are not only powerful and generalizable but also inherently explainable, self-consistent, and grounded in the most fundamental principles of the universe.

\bibliographystyle{plain}

\end{document}